\documentclass[prd,floatfix,showpacs]{revtex4-1}
\usepackage{amsmath}
\usepackage{amsfonts}
\usepackage{amssymb}
\usepackage{graphicx}
\usepackage[margin=3.0cm]{geometry}
\usepackage{hyperref}

\newcommand\beq{\begin{equation}}      
\newcommand\beqnn{\begin{eqnarray*}}   
\newcommand\beqa{\begin{eqnarray}}     
\newcommand\beqann{\begin{eqnarray*}}  

\newcommand\eeq{\end{equation}}        
\newcommand\eeqnn{\end{eqnarray*}}     
\newcommand\eeqa{\end{eqnarray}}       
\newcommand\eeqann{\end{eqnarray*}}    

\newcommand\bi{\begin{itemize}}
\newcommand\ei{\end{itemize}}

\begin{document}

\title{A unifying perspective on the Moyal and Voros products and their physical meanings}
\date{\today}
\author{Prasad Basu$^{a}$, Biswajit Chakraborty$^{a,c}$ and \footnote{Corresponding author: fgs@sun.ac.za}Frederik G Scholtz$^{b,c}$}
\affiliation{$^a$S.~N.~Bose National Centre for Basic Sciences,JD Block, Sector III, Salt Lake, Kolkata-700098, India\\$^b$National Institute for Theoretical Physics (NITheP),
Stellenbosch 7600, South Africa\\
$^c$Institute of Theoretical Physics,
University of Stellenbosch, Stellenbosch 7600, South Africa}

\begin{abstract}
\noindent
The Moyal and Voros formulations of non-commutative quantum field theory has been a point of controversy in the recent past.  Here we address this issue in the context of non-commutative non-relativistic quantum mechanics.  In particular we show that the two formulations simply correspond to two different representations associated with two different choices of basis on the quantum Hilbert space.  From a mathematical perspective the two formulations are therefore completely equivalent, but we also argue that only the Voros formulaton admits a consistent physical interpretation. These considerations are elucidated by considering the free particle transition amplitude in the two representations.

\end{abstract}

\maketitle

There is a recent upsurge of studies in non-commutative quantum theories and this is based 
on the belief that this can provide another window into the nature of space-time at Planck 
scale physics and perhaps can complement the insights gained through other approaches like 
string theory and loop quantum gravity (see e.g. \cite{confser} for a recent overview).  Aside from this non-commutativity has relevance for condensed matter phenomena like the quantum Hall effect \cite{bell} and topological insulators \cite{prod}. 

In their seminal paper, Doplicher et al. \cite{dop} argued from the considerations of both general 
relativity and quantum mechanics that the localization of an event in space-time with 
arbitrary accuracy is operationally impossible.  This feature is captured by postulating a non-vanishing commutation relation between the 
coordinates, which are now elevated to the level of operators. The simplest form of 
non-commutative space-time takes the form of 
\begin{eqnarray}
[\hat x^{\mu}, \hat x^{\nu}]=i\theta^{\mu \nu},
\label{theta}
\end{eqnarray}
where $\Theta=\{\theta^{\mu \nu}\}$ is a constant matrix and is usually taken to be same in 
all coordinate systems just like other fundamental constants such as $\hbar$ and $c$, {\it i.e.} they  do not transform like a second rank antisymmetric 
tensor \cite{asch}. Most studies of quantum theories defined on a non-commutative space-time is 
based on this simplest form of non-commutativity. In the conventional method of analyzing 
quantum field theories on such non-commutative space-times one usually demotes the above 
mentioned operator-valued coordinates to c-number valued coordinates, but now composing 
through appropriately defined star-products. Out of many choices of the star-products, 
the two most popular ones are the Moyal and Voros star-product. The respective star-brackets 
yield coordinate algebras which are isomorphic to (\ref{theta}) in both the cases 
\begin{eqnarray}
[x^{\mu}, x^{\nu}]_{\star_{M/V}}=x^{\mu}\star_{M/V}x^{\nu}-x^{\nu}\star_{M/V}x^{\mu}=i\theta^{\mu \nu},
\end{eqnarray}
where the $\star_M$ and $\star_V$ denotes the Moyal and the Voros star products, respectively. 
In 2+1 dimension they are defined as 
\begin{eqnarray}
f(\vec x)\star_{M/V} g(\vec x)=(f\star_{M/V} g)(\vec x)=f(\vec x)\exp\Bigl(\frac{i}{2}\theta^{i j (M/V)}\overleftarrow \partial_{i} \overrightarrow \partial_{j} \Bigr)g(\vec x),
\label{strprdcts}
\end{eqnarray}
where the matrices $\Theta^{(M/V)}=\{\theta^{ij (M/V)}\}$ are given by
\begin{eqnarray}
\Theta^{M}=\begin{pmatrix}
0 & 0 & 0\\
0 & 0 & \theta \\
0 & -\theta & 0
\end{pmatrix},  \ \ \ \ \ \  
\Theta^{V}=\begin{pmatrix}
0 & 0 & 0\\
0 & -i\theta & \theta \\
0 & -\theta & -i\theta
\end{pmatrix},
\label{thetamatrx}
\end{eqnarray} 
if the time is taken to be an ordinary $c$-number.  There is a certain equivalence between these two star-products, for example there exists an equivalence map $T=e^{\frac{\theta}{4}\nabla^2}$ connecting Moyal and Voros star-products: 
\begin{eqnarray}
T(f\star_M g)(x)=(T(f)\star_V T(g))(x),
\label{tmap}
\end{eqnarray}
which makes the free theories defined in terms of these two star-products essentially equivalent. There are, however, controversies regarding the persistence of this equivalence at the level of interacting quantum field theories \cite{bal1,lizzi}. 

In this paper we address these issues at the level of non-commutative non-relativistic quantum 
mechanics.  The main motivation for undertaking such an investigation stems from the fact that a non-relativistic limit of any quantum field theory having massive quanta (like Klein-Gordon and Dirac fields) goes 
over to a non-relativistic Schr\"odinger field theory in its second quantized version, whereas, 
the corresponding first quantized version corresponds to non-relativistic quantum mechanics. 
One therefore expects that these different choices of star products that are made at the level of the non-commutative field theory must also manifest itself in some or other way on the level of the underlying non-commutative non-relativistic quantum system.  In other words, we expect that the field theoretical models studied in 
\cite{bal1,lizzi} should admit a consistent quantum mechanical description in their non-relativistic limits where the mathematical and, particularly the physical, manifestations of these different choices of star products may be clearer. 

We organize the paper by first giving a brief review of the operator approach to non-commutative quantum mechanics, which does not depend on the choice of any particular star-product. Next we demonstrate how the Moyal and Voros formulations of non-commutative quantum mechanics arise simply as different, but equivalent, representations corresponding to different choices of bases, which we'll refer to as the Moyal and Voros bases. We then proceed to discuss the physical interpretations of these two bases.  Finally we discuss how these two representations manifest themselves on the level of transition amplitudes.            
    
It is important, as was already emphasized in the literature \cite{lizzi}, that there is a completely general and abstract operator formulation of non-commutative quantum field theory and quantum mechanics.  This approach to non-commutative quantum mechanics was advocated in \cite{BC1} and the advantage of this formulation to solve non-trivial systems such as a spherical well, which is virtually intractable in a star product formulation involving infinite order derivatives, was demonstrated in \cite{BC2}. In this setting one would expect the Moyal and Voros formulations to arise from different 'position' representation as in ordinary quantum mechanics.  To fix notations and sketch the necessary background we briefly review this formalism.
    
In $2+1$ dimension the matrix $\Theta$ of eq. (\ref{theta}) takes the form:
\begin{eqnarray}
\Theta=
\begin{pmatrix}
0 & 0 & 0 \\
0 & 0 & \theta \\
0 & -\theta & 0
\end{pmatrix},
\label{thetamatrix}
\end{eqnarray}
if the time $t$ is taken to be commuting so that there is no spatio-temporal non-commutativity. 
Note from eq. (\ref{thetamatrx}) that $\Theta^{M}=\Theta$, but that $\Theta^V\neq \Theta$,\  as \ $\Theta^{V}$ contains 
imaginary diagonal entries. This has important bearings on our subsequent analysis, as we shall see.  
The corresponding non-commutative Heisenberg algebra can be written as (we work in units with $\hbar=1$)
\begin{eqnarray}
\label{heiss}
\left[\hat x_i, \hat x_j\right]&=&i\theta \epsilon_{ij},\nonumber\\
\left[\hat x_i, \hat p_j\right]&=&i\delta_{ij},\\
\left[\hat p_i, \hat p_j\right]&=&0.
\end{eqnarray}

One can construct standard creation and annihilation operators $b^{\dagger}$ and $b$:
\begin{eqnarray}
b=\frac{\hat x_1+i \hat x_2}{\sqrt{2\theta}}, \ \ \ \ \ b^{\dagger}=\frac{\hat x_1-i \hat x_2}{\sqrt{2\theta}}.
\label{bbdagger}
\end{eqnarray} 
The non-commutative plane can therefore be viewed as a boson Fock space spanned by the eigenstate $|n\rangle$ of the operator 
$b^{\dagger}b$. We refer to it as the classical configuration space (${\mathcal H}_c$):
\begin{eqnarray}
{\mathcal H}_c&=&span \Bigl\{ |n\rangle=\frac{1}{\sqrt{n!}}(b^{\dagger})^n|0\rangle\Bigr\}^{n=\infty}_{n=0}   
\end{eqnarray}
Note that this space plays the same role as the classical configuration space ${\mathcal R^2}$ in commutative quantum mechanics.  Next we introduce the quantum Hilbert space in which the states of the system and the non-commutative Heisenberg algebra are to be represented. This is taken to be the set of all bounded trace-class operators (the Hilbert-Schmidt operators) over ${\mathcal H}_c$ and we refer to it as quantum Hilbert space,(${\mathcal H}_q$),
\begin{eqnarray}
{\mathcal H}_q=\{ \psi(\hat x_1, \hat x_2): {\rm tr_c}(\psi^{\dagger}\psi)<\infty\}.
\end{eqnarray}
Physical states are represented by the elements of ${\mathcal H}_q$ and are denoted by round brackets
$\psi(\hat x_1, \hat x_2)\equiv |\psi)$.
The inner product is defined as
\begin{eqnarray}
(\phi|\psi)={\rm tr_c}(\phi^{\dagger}\psi),
\label{inprdct}
\end{eqnarray}
where the subscript $c$ refers to tracing over ${\mathcal H}_c$.  If $\hat X_i$, $\hat P_i$ are the representations of the operators 
$\hat x_i$ and $\hat p_i$ acting on ${\mathcal H}_q$, then a unitary representation is obtained by the following action:
\begin{eqnarray}
\hat X_i \psi(\hat x_1, \hat x_2)=\hat x_i \psi(\hat x_1, \hat x_2),\nonumber \\
\hat P_i \psi(\hat x_1, \hat x_2)=\frac{1}{\theta}\epsilon_{ij}[\hat x_j, \psi(\hat x_1, \hat x_2)].
\label{ncrep}
\end{eqnarray} 
It is easily verified that the momentum eigenstates $|p)$ are given by
\begin{eqnarray}
|p)=\sqrt{\frac{\theta}{2\pi}}e^{ip\cdot \hat x},\ \ \ \ \hat P_i |p)=p_i |p), 
\label{peigenst}
\end{eqnarray}
and that they satisfy the usual resolution of identity and orthogonality condition
\begin{eqnarray}
 \int d^2p |p)(p|&=& {\bf 1}, \ \ \ \ \ (p|p^\prime)=\delta^2(p -p^\prime).
 \label{pcomplt}
\end{eqnarray}

Following the analogy of coherent states of the Harmonic oscillator, one can introduce minimum uncertainty states in the classical configuration space  
\begin{eqnarray}
|z\rangle=e^{-\bar zb+zb^{\dagger}}|0\rangle=e^{-\frac{1}{2}|z|^2}e^{zb^\dagger}|0\rangle \in \mathcal{H}_{c} 
\label{cstateHc}
\end{eqnarray}
satisfying 
\begin{eqnarray}
b|z\rangle=z|z\rangle
\label{cstateegnv}
\end{eqnarray}
for an arbitrary complex number $z$. From this a basis $|z,\bar z)=|z\rangle\langle z|\in {\mathcal H}_q$ can be constructed for the quantum Hilbert space.  In particular they satisfy  
\begin{eqnarray}
B|z, \bar z)&=&z|z,\bar z),\nonumber\\
(z^\prime, \bar z^\prime|z, \bar z)&=&tr_{c}[(|z^\prime\rangle\langle z^\prime|)^{\dagger}(|z\rangle\langle z|)]=e^{-|z-z^\prime|^2}
\label{inpro}
\end{eqnarray}
and, most importantly, the completeness relation \cite{BC1}
\begin{equation}
\int \frac{d^2z}{\pi}|z,\bar z)\star_V(z, \bar z|=\mathbf{1}.
\label{vcomplt}
\end{equation}
Here $B=\frac{\hat X_1+i\hat X_2}{\sqrt{2\theta}}$ is the 
representation of the operator $b$ on ${\mathcal H}_q$ and 
the Voros-star product $\star_V$ defined in (\ref{strprdcts})
takes the form
\begin{equation}
f(z, \bar z)\star_V g(z, \bar z)=f(z, \bar z)e^{\overleftarrow \partial_z \overrightarrow \partial_{\bar z}} g(z, \bar z).
\label{vstar}
\end{equation} 
We refer to this basis as the Voros basis. 
The overlap of this basis with the momentum eigenstate (\ref{peigenst}) 
is given by
\begin{eqnarray}
(z,\bar z|p)&=&\sqrt{\frac{\theta}{2\pi}}e^{-\frac{\theta |p|^2}{4}} 
e^{i\sqrt{\frac{\theta}{2}}(p\bar z+\bar p z)} \nonumber \\
&=&\sqrt{\frac{\theta}{2\pi}}e^{-\frac{\theta |p|^2}{4}}e^{ip\cdot x}, 
\label{pzoverlap}
\end{eqnarray} 
where we have introduced the Cartesian coordinates 
\begin{eqnarray}
x_1=\sqrt{\frac{\theta}{2}}(z +\bar z) \ \ \ \ \text{and}  \ \ \ \  x_2=i \sqrt{\frac{\theta}{2}}(\bar z - z),
\label{zxreln}
\end{eqnarray}  
so that these Voros states can alternatively be labelled as $|x)_V\equiv |z, \bar z)$.
From this we infer that we may expand the Voros basis states as follow in terms of momentum states
\begin{eqnarray}
|x)_V=\sqrt{\frac{\theta}{2\pi}}\int d^2p e^{-\frac{\theta p^2}{4}}e^{-ip\cdot x}|p)=\int\frac{d^2p\theta}{2\pi}e^{-\frac{\theta p^2}{4}}e^{ip\cdot(\hat x-x)}.
\end{eqnarray}

Next we introduce what we refer to as the Moyal basis, defined as an expansion in terms of momentum states as follow
\begin{equation}
\label{pxpanson}
|x)_M=\int \frac{d^2p}{2\pi}e^{-ip\cdot x}|p)=\sqrt{\frac{\theta}{2\pi}}\int\frac{d^2p}{2\pi}e^{ip\cdot(\hat x-x)}.
\end{equation}
These states satisfy
\begin{eqnarray}
\int d^2x |x)_M\star_M{}_M(x|&=&\int d^2x|x)_M{}_M(x|={\bf 1},\nonumber\\
\label{mcomplt1}
(p|x)_M&=&\frac{1}{2\pi}e^{-ip\cdot x},\nonumber\\
{}_M(x|x^\prime)_M&=&\delta^2(x-x^\prime ).
\label{mortho}
\end{eqnarray} 
The Moyal basis is therefore an orthogonal basis, unlike the Voros basis. Using eqs.(\ref{pxpanson}), (\ref{pzoverlap}) and (\ref{mortho}) we find the overlap between the Moyal and Voros basis vectors to be 
\begin{equation}
{}_V(x^\prime|x)_M=\sqrt{\frac{2}{\pi \theta}}e^{-\frac{(x-x^\prime)^2}{\theta}}.
\label{QMVprodct}
\end{equation} 
In ${\mathcal H}_q$ one can define operators $\hat X^c_i$ as \cite{bal3}
\begin{eqnarray}
\hat X^c_i=\hat X_i+\frac{\theta}{2}\epsilon_{ij}\hat P_j.
\label{XC}
\end{eqnarray} 
These operators are mutually commuting {\it i.e.} $[X^c_i,X^c_j]=0$ 
and become identical to the position operator in $\theta=0$ case. 
An interesting property of the $\hat X_i^c$'s is that they are diagonal
in the Moyal basis: 
\begin{eqnarray}
{}_M(x^\prime|\hat X^c_i|x)_M&=&{}_M(x^\prime|\hat X_i|x)_M+\frac{\theta_{ij}}{2}(x^\prime|\hat P_j|x) \nonumber \\
&=& {}_M(x^\prime|\hat X_i|x)_M+ \frac{\theta_{ij}}{8\pi^2}\int d^2p\, p_je^{ip\cdot(x^\prime -x)}.
\end{eqnarray} 
Using (\ref{bbdagger}), (\ref{inprdct}), (\ref{peigenst}) and (\ref{pxpanson}) one can prove that 
\begin{eqnarray}
{}_M(x^\prime|\hat X_i|x)_M=x_i \delta(x -x^\prime) -\frac{\theta_{ij}}{8\pi^2}\int d^2p\, p_je^{ip\cdot(x^\prime-x)}.
\label{xcdiagnl}
\end{eqnarray} 
Therefore,
\begin{eqnarray}
{}_M(x^\prime|\hat X^c_i|x)_M=x_i \delta(x -x^\prime),
\label{xcdiagnl1}
\end{eqnarray}
which implies 
\begin{eqnarray}
\hat X^c_i|x)_M=x_i |x)_M, 
\end{eqnarray}
{\it i.e.}, the Moyal basis states are simultaneous eigenstate of the operators $X^c_i$. 

We can impose the additional structure of an algebra on the quantum Hilbert space by defining the multiplication map $\mu:{\mathcal H}_q\otimes {\mathcal H}_q\rightarrow {\mathcal H}_q$ as follows 
\begin{equation}
\label{product}
\mu\left(|\psi)\otimes |\phi)\right)=|\psi\phi)
\end{equation}
where ordinary operator multiplication is implied on the right.  

We can now pose the following question, what is the form of the representation of this product state when represented in the Moyal or Voros basis and, in particular, is there a composition rule in terms of the representations of the individual states in these bases.  To calculate this, we expand an arbitrary state $|\psi)$, in analogy of eq.(\ref{pxpanson}), in terms of the momentum basis as follows:
\begin{equation}
|\psi)=\sqrt{\frac{\theta}{2\pi}}\int\frac{d^2p}{2\pi}\psi(p)e^{ip\cdot \hat x}.
\end{equation} 
Note that the condition of normalizability of the state, i.e., $(\psi|\psi)={\rm tr_c}(\psi^\dagger\psi)<\infty$ implies that the function $\psi(p)$ must be square integrable.

The computation is now straightforward and one finds
\begin{eqnarray}
\label{map1}
{}_M(x|\psi\phi)&=&\sqrt{2\pi\theta} {}_M(x|\psi)\star_M{}_M(x|\phi),\\
\label{map2}
{}_V(x|\psi\phi)&=&4\pi^2{}_V(x|\psi)\star_V{}_V(x|\phi),
\end{eqnarray}
where 
\begin{eqnarray}
\label{moywf}
{}_M(x|\psi)&=&\int\frac{d^2p}{(2\pi)^2}\psi(p)e^{ip\cdot x},\\
\label{vorwf}
{}_V(x|\psi)&=&\sqrt{\frac{\theta}{2\pi}}\int\frac{d^2p}{(2\pi)^2}\psi(p)e^{-\frac{\theta p^2}{4}}e^{ip\cdot x}=\sqrt{\frac{\theta}{2\pi}}e^{\frac{\theta \nabla^2}{4}}{}_M(x|\psi)
\end{eqnarray}

The unimportant prefactors on the right depend on the normalisation convention of the basis states (note that it is dimensionful in the Moyal basis as the wave functions have the dimensions of an inverse length).  The important message of this result is that the Moyal and Voros compositions are simply related to the choice of basis made when a 'position representation' of the quantum Hilbert space is constructed. In the Voros basis, related to states of minimum position uncertainty, the Voros composition rule results and in the case of the Moyal basis, which are eigenstates of the commuting operators $X_i^c$, one finds the Moyal composition rule.  Mathematically these two representations are, of course, completely equivalent. There are, however, a number of subtleties to note.  In the Moyal basis one has to take the wave function to belong to the set of Schwartz class function in order to ensure that the functions and all its derivatives vanish fast enough at infinity to avoid unwanted boundary terms. Without this the first equation in (\ref{mortho}) will not hold.  The same consideration applies to the Voros basis, but there is a further consideration that needs to be taken into account in this case.  The point to note, which is often not appreciated in the literature, is that the $T$ operator $T=e^{\frac{\theta \nabla^2}{4}}$ relating the Moyal and Voros basis is not invertible on the space of square integrable or even Schwartz class functions.  To see this consider a Gaussian function $f(x)=e^{-\frac{x^2}{\alpha^2}}$.  Taking the Fourier transform and applying the inverse operator $T^{-1}=e^{-\frac{\theta \nabla^2}{4}}$ gives
\begin{equation}
T^{-1}f(x)=\frac{\alpha^2}{4\pi}\int d^2p e^{-\frac{1}{4}(\alpha^2-\theta) p^2+ip\cdot x}.
\end{equation}
This integral clearly does not exist when $\alpha<\sqrt{\theta}$.  This shows that although $f(x)$ is of Schwartz class, $T^{-1}f(x)$ does not exist if the function $f(x)$ varies too rapidly on length scales of order $\sqrt{\theta}$.  The operator $T^{-1}$ is only defined on the class of functions appearing on the left of eq. (\ref{vorwf}), with $\psi(p)$ square integrable as required by the finite norm of $|\psi)$.  

The above considerations imply some care in interpreting the equivalence between the Moyal and Voros representations.  Clearly this is not an equivalence of two quantum systems defined on spaces of square integrable functions.  The Moyal representation is defined on the space of functions that are of Schwartz class in configuration space, while the Voros representation is defined on a smaller subspace that also requires smoothness on small length scales as reflected in (\ref{vorwf}) through the suppression of high momentum modes.  This is not unexpected as the Hilbert space of functions in the Voros representation cannot include functions that violate the space-space uncertainty relations, e.g., functions such as the Gaussians above with spread $\alpha<\sqrt{\theta}$. These considerations may also play a role in the two different representations of a quantum field theory as the classes of functions over which are integrated in the path integral are different in the two cases.  Whether this plays a role in the differences reported in \cite{bal1} between quantum field theories on the Moyal and Voros planes needs further investigation. 

There are two more points to note.  Firstly, the $T$ operator is not unitary from which the non-orthogonality of the Voros states stems.  Secondly note that (\ref{map2}) encodes the map of (\ref{tmap}).  This is easily seen by substituting the second equality of (\ref{vorwf}) into (\ref{map2}) and using (\ref{map1}) on the left. Note, though, that the functions $f$ and $g$ of (\ref{tmap}) have to be identified with the Moyal wave functions on either side, although it is the Voros product that appears on the right.  

Despite the mathematical equivalence of these two representations, one needs to take care with the physical interpretation of these two representations, as was already pointed out in \cite{lizzi} and on which we further elaborate now.  

Let us consider the physical interpretation of the Moyal and Voros bases.  What we have in mind is the question whether the system can be prepared in these states, which is closely related to the issue of measurement and the status of these operators as physical observables.  It should be immediately clear that the commuting operators $X_i^c$, of which the Moyal states are eigenstates, cannot play the role of physical position observables as this will violate the minimum uncertainty in position deriving from non-commutativity.  These operators must therefore either be unobservable, in which case we cannot prepare the system in a Moyal state, or else the physical meaning of the quantum number $x$ must be different from physical position \cite{sch}.  Either way we must accept that we cannot interpret the Moyal states, at least not in the context of strong measurements based on projective valued measures (PVM's), as representing a system spatially localized at position $x$, they are either purely mathematical constructs or at best they represent a physical state which does not correspond to spatial localization.

In contrast to the Moyal basis the Voros basis does not violate the minimum uncertainty in position.  Indeed, they represent the optimal spatial localization and therefore one would expect them to allow the standard quantum mechanical interpretation of weak measurements.  This is indeed the case as has been elaborated on elsewhere \cite{BC1,sch} and we only briefly recap the essentials of these discussions here.  From the non-commutative nature of space it is clear that the conventional interpretation of strong measurements based on projective valued measures cannot apply.  Instead one must think about position measurement in the sense of a weak measurement based on a positive operator valued measure (POVM's) \cite{ber}.  Indeed one can easily verify that the operators
\begin{equation}
\pi_x=\frac{1}{\pi}|x)_V\star_V{}_V(x|
\end{equation}             
form a positive operator valued measure, i.e., they are positive and integrate to the identity from (\ref{vcomplt}).  The only difference from a standard PVM is the non-orthogonality of these operators, which requires a relaxation of von Neuman's projective assumption and changes the measurement into a weak measurement.  A completely consistent probability interpretation now follows and the probability of finding the system at the point $x$ under a position measurement, given that the system is described by a density matrix $\rho$, is simply
\begin{equation}
P(x)={\rm tr}_q (\pi_x\rho),
\end{equation}
where the subscript $q$ denotes the trace over quantum Hilbert space.  For a pure state $\rho=|\psi)(\psi|$ this reduces to  
\begin{equation}
P(x)=\langle x|\psi|x \rangle\star_V\langle x|\psi|x \rangle,
\end{equation}
where $|x\rangle$ is the coherent state of eq. (\ref{cstateHc}).  This closely resembles our standard position probability interpretation and, indeed, reduces to it in the commutative limit \cite{BC1}. From this discussion it should be clear that the Voros, rather than the Moyal, basis represents a system maximally localized at a point $x$ and that these states are therefore the appropriate physical states describing such a system. A final point to note is that one cannot introduce a POVM based on the Moyal product as this will not satisfy the positivity condition.  This again hints at the Voros product, and thus Voros basis, as the appropriate physical framework. 

We have now established that, although the Moyal and Voros bases are mathematically completely equivalent, only the Voros basis can serve as the physical framework to describe a maximally localized particle.  This also has profound implications for other physical quantities such as transition amplitudes. These will clearly be different when represented in the Moyal and Voros basis, yet only the transition amplitudes between Voros basis states have physical meaning.  Great care should therefore be exercised in the physical interpretation of a Moyal or Voros basis representation, as was already pointed out in \cite{lizzi} where similar conclusions were drawn in the context of non-commutative relativistic quantum field theory.

To clarify these remarks in the present setting, we briefly discuss the transition amplitudes in the Moyal and Voros representations.  The transition amplitude for the Voros basis was already calculated trough a path integral formalism in \cite{sunandan}.  Here we briefly repeat this calculation for a free particle in the Moyal basis. The free particle kernel is given by
\begin{eqnarray}
K(x_f,t_f;x_i, t_i)&=&{}_M(x_f, t_f|x_i, t_i)_M \nonumber \\
&=&{}_M( x_f|e^{-i(t_f-t_i)\hat H}| x_i)_M,
\label{krnl1}
\end{eqnarray} 
where, $\hat H=\frac{\hat P^2}{2m}$ is the free particle Hamiltonian in $\mathcal{H}_q$ and 
$| x,t)_M$ is the Heisenberg basis corresponding to $| x)_M$ at time $t$ defined as 
\begin{eqnarray}
| x,t)_M=e^{i\hat H t}|x)_M.
\end{eqnarray} 
They also satisfy the completeness relation
\begin{eqnarray}
\int d^2x| x,t)_M\star_M {}_M( x,t|=\int d^2x|x,t)_M{}_M( x,t|={\bf 1}.
\end{eqnarray} 
We divide the time interval into $N+1$ equal intervals corresponding to $t_i$, $i=1, N$ such that 
$t_{i+1}-t_i=\epsilon\,,\forall i$.  Inserting the complete set $|x_i, t_i)$ 
for each $t_i$ in the kernel we get
\begin{eqnarray}
K(x_f,t_f; x_i, t_i)&=&{}_M(x_f, t_f|x_i, t_i)_M \nonumber \\
&=&\int .. \int d^2x_N .. d^2x_1 {}_M(x_f, t_f|x_N, t_N)_M{}_M(x_N,t_N|x_{N-1},t_{N-1})_M..{}_M(x_1,t_1|x_i, t_i)_M.\nonumber\\
\label{krnl2}
\end{eqnarray}
The kernel $(\vec x_{i+1}, t_{i+1}f|x_i, t_i)$ corresponding to the infinitesimal time interval $\epsilon=t_{i+1}-t_i$ 
takes exactly the same form as the corresponding kernel in the commutative case. Therefore $K(x_f,t_f;x_i, t_i)$ is given by the commutative kernel
\begin{eqnarray}
K(x_f,t_f;x_i, t_i)&=&{}_M(x_f, t_f|x_i, t_i)_M\nonumber\\
&=&\Bigl[\frac{m}{2\pi iT} \Bigr]\exp\Bigl[\frac{im(x_f -x_i)^2}{2T}\Bigr], 
\end{eqnarray} 
where $T=t_f-t_i$.  

The transition amplitude from a voros-state $|x_i, t_i)_V$ at time $t_i$ to another voros-state $|x_f, t_f)_V$ at time $t_f$
is related to the corresponding Moyal kernel $(x_f, t_f|x_i, t_i)$ as
\begin{eqnarray}
 {}_V( x_f,t_f|x_i, t_i)_V&=&\int d^2x_f^\prime d^2x_i^\prime {}_V( x_f,t_f|x_f^\prime, t_f)_M{}_M(x_f^\prime,t_f|x_i^\prime, t_i)_M{}_M(x_i^\prime, t_i|x_i, t_i)_V 
\nonumber \\
 &=&\int d^2x_f^\prime d^2x_i^\prime {}_V(x_f|x_f^\prime)_M K(x_f^\prime, t_f;x_i^\prime, t_i){}_M(x_i^\prime|x_i)_V.
\end{eqnarray} 
This is an important relation relating the kernels in Voros and Moyal basis and the former is obtained from the latter 
by convoluting w.r.t the overlap of these two bases given in (\ref{QMVprodct}) at initial and final positions. After a 
straightforward computation, this yields     
\begin{eqnarray}
\label{vtran}
{}_V(x_f, t_f|x_i, t_i)_V=\frac{m}{(m\theta+iT)}exp\Bigl[-\frac{m(x_f-x_i)^2}{2(m\theta +iT)}\Bigr],
\end{eqnarray} 
which agrees with the result of \cite{sunandan}. This demonstrates the remarks made above, namely, that even the free particle 
transition amplitudes are different, which stems from the different choices of basis. The Moyal transition amplitude can, however, not be given the physical interpretation of propagation from $x_i$ to $x_f$.  This is given by the Voros amplitude, which, not surprisingly, involves an averaging over the initial and final Moyal states, reflecting the fuzziness inherent to the Voros basis and non-commutative space which is not captured in the Moyal basis. When a potential or interactions are included matters become even more subtle as the choice of coordinates ($\hat X_i$ or $\hat X_i^c$) involved in the potential or interaction also lead to physical observable effects.     

The same conclusions can also be reached from a Moyal or Voros field theoretic formulation of the non-relativistic Schr\"odinger equation.  The free particle action in the commutative case is given by
\begin{eqnarray}
S_c=\int(i\bar \psi \frac{\partial\psi}{\partial t}+\frac{1}{2m}\bar \psi \nabla^2\psi)d^2xdt.
\end{eqnarray}
The corresponding Moyal and Voros actions in non-commutative space are given by 
\begin{eqnarray}
S_{(M/V)}=\int(i\bar \psi \star_{M/V} \frac{\partial\psi}{\partial t}+\frac{1}{2m}\bar \psi\star_{M/V} \nabla^2\psi)d^2xdt.
\end{eqnarray}
From the properties of the Moyal product one easily establishes that $S_M=S_c$, but $S_V$ 
is different from $S_c$. 

Varying with respect to $\bar\psi$ gives the equation of motion:
\begin{eqnarray} 
e^{-\frac{\theta}{2}\nabla^2}[i\frac{\partial }{\partial t}+\frac{1}{2m}\nabla^2]\psi=0.
\end{eqnarray}
The free particle kernel $K(x_f,t_f;x_i, t_i)$ satisfies the equation
\begin{eqnarray}
e^{-\frac{\theta}{2}\nabla_f^2}[i\frac{\partial }{\partial t}+\frac{1}{2m}
\nabla^2_f]K(x_f,t_f;x_i, t_i) \nonumber \\
=\delta^2(x_f -x_i)\delta(t_f-t_i).
\end{eqnarray}
The momentum space Green's function (Fourier transform of $K(x_f,t_f;x_i, t_i)$) is given by
\begin{eqnarray}
G(k, \omega_k )=\frac{1}{(2\pi)^3}\frac{e^{-\frac{\theta k^2}{2}}}{(\omega_k-\frac{k^2}{2m})}.
\end{eqnarray}
The retarded Green's function $K(x_f,t_f;x_i, t_i)$ is calculated from the knowledge of 
$G(k, \omega_k)$ by taking care of the poles in the usual manner: 
\begin{eqnarray} 
K(x_f,t_f;x_i, t_i)&=&\int G(k, \omega_k)e^{-i[\omega_k(t_f-t_i)- k\cdot(x_f-x_i)}d^2kd\omega_k \nonumber \\
&=&\frac{m}{[m\theta+iT]}exp\Bigl[-\frac{m(\vec x_f-\vec x_i)^2}{2(m\theta +iT)}\Bigr],
\end{eqnarray}
which again agrees with (\ref{vtran}) and the result of \cite{sunandan}.   

To conclude, we have shown that the Moyal and Voros formulations correspond to two different representations associated with two different choices of basis in the quantum Hilbert space.  Mathematically they are therefore completely equivalent.  However, we have also argued that not both representations admit a consistent physical interpretation and that only the Voros states can be interpreted as describing a maximally localized system.  This is also reflected in the transition amplitudes that differ with only the Voros amplitude representing the physical transition amplitude. Also note that when one computes spatial correlation functions they are basis dependent and will yield different results in the Moyal and Voros bases as was already pointed out in \cite{sunandan1, sunandan2}.  Again the Voros based correlation functions correspond to physical spatial correlations. S-matrix elements are computed between asymptotic free particle states, which are eigenstates of momentum and, since momenta commute, there is no ambiguity in these states.  Since the Moyal and Voros bases are merely choices of basis, which are mathematically equivalent, one expects the Voros and Moyal formulations to yield the same S-matrices.  This was indeed confirmed in \cite{lizzi}.  Although these results concur with those of \cite{lizzi}, different conclusions were drawn in \cite{bal1}.  These stem from the different approaches between \cite{lizzi} and \cite{bal1} in the construction of the quantum field theory.  In \cite{bal1} it is argued that physical differences arise between quantum field theories on the Moyal and Voros plane and that the Moyal formulation is favoured by consistency.  As is well known the arguments presented here in the context of ordinary quantum mechanics with a finite number of degrees of freedom do not necessarily extend to the quantum field theory with an infinite number of degrees of freedom, e.g, anomalies may arise.  In addition position does not directly correspond to any observable in Quantum Field Theory, which may have some conceptual implications.
This again emphasizes the need for a careful analysis of these different twistings that realize the spatial non-commutativity and perhaps even the necessity of treating the non-commutative quantum field theory exclusively at the abstract operator level which does not invoke any star product.  

{\it Acknowledgements:}  Support under the Indo-South African research agreement between the Department of Science and Technology, Government of India and the National Research Foundation of South Africa is acknowledged, as well as a grant from the National Research Foundation of South Africa.

\end{document}